# Cylindrical quantum wires with hydrogen-bonded materials

**Vjekoslav Sajfert[†], Rajka Đajić[‡], Željka Cvejić[‡]**

**Abstract**

Properties of cylindrical quantum wires are analysed in this paper. Energies of elementary excitations as well as one-particle wave functions were found for mentioned structure. For cylindrical quantum wires the temperature of phase transition was found. The behaviour of electric susceptibility in paraelectric phase was investigated.

[†] Technical Faculty "Mihajlo Pupin", Zrenjanin
[‡] Institute of Physics, Faculty of Science, Novi Sad

# Cylindrical quantum wires with hydrogen-bonded materials

**Introduction**

Cylindrical structures are often used in practice (electrical conductor, water tubes). On the other hand these structures appear in animated matter (nerves, blood vessels etc.). Therefore the permanent interest in investigations of these structures is understandable.

The goals of our investigation are micro and nano cylindrical structures. The basic characteristics of cylindrical quantum wires will be analysed. The length of quantum wires will be considered as sufficiently big to make boundary conditions negligible and therefore those will be treated as translatory invariant along their axis. In hollow quantum tubes molecules are located in shell of cylinder. One cross-section of the tube contains relatively small number of molecules, which is of the order 5-15 molecules.

The main characteristic of cylindrical structures is cyclical invariance in discs. The consequences of this invariance are series of interesting effects, which are sometimes of practical interest.

In the first part of this paper the Hamiltonian of hydrogen-bonded quantum wires will be formulated.
 and will be analysed.

In the second part dynamical as well as thermodynamical properties will be investigated.

**1. Hydrogen-bonded quantum tubes**

As it was said in Introduction hydrogen-bonded quantum tubes are microscopically long cylinders with microscopic cross-sections. Hydrogen-bonds are located in shell of the cylinder.

In analyses the nearest neighbours' approximation will be used. In this approximation every bond has two neighbours' along the axis of cylinder and two neighbours' at the cycle which represents cross-section of cylinder.

In analyses of hydrogen-bonded systems the quasi-spin idea is used. One half spin is corresponded to two possible positions of proton in W potential [1-3]. This potential is a consequence of interaction of two negative ions of oxygen or some close elements (N, C etc.). Discrete changes of proton position are corresponded to two projections of spin $S=1/2$. Since the spin is $S=1/2$, its components can be represented through Pauli operators' [4], whose commutation relations are:

$$[P_n, P_m^+] = (1 - 2 P_n^+ P_n)\delta_{n,m}; [P_n, P_m] = [P_n^+, P_m^+] = 0 \quad (1.1)$$

$$P_n^{+2} = P_n^2 = 0$$

In analyses will be used Hamiltonian of the Ising model in transferse field (IMTF [5]) which is unitary transformed so that the third order processes are included in quadratic part [6].

Taking this into account, the Hamiltonian of this system, taken in the nearest neighbors' approximation, can be written as follows:

$$H = \frac{1}{2}\sum_{\vec{m},\vec{n}} \left( J_{m,n;m+1,n} + J_{m,n;m-1,n} + I_{m,n;m,n+1} + I_{m,n;m,n-1} \right) P_{m,n}^+ P_{m,n} -$$

$$-\frac{1}{2}\sum_{n\nu}P^+_{m,n}\left(Y_{m,n;m+1,n}P_{m+1,n}+Y_{m,n;m-1,n}P_{m-1,n}+W_{nm,n;m,n+1}P_{m,n+1}+W_{m,n;m,n-1}P_{m,n-1}\right)-$$

$$-\frac{1}{2}\sum_{n\nu}P^+_{m,n}P_{m,n}(J_{m,n;m+1,n}P^+_{m+1,n}P_{m+1,n}+J_{m,n;m-1,n}P^+_{m-1,n}P_{m-1,n}+$$

$$+I_{m,n;m,n+1}P^+_{m,n+1}P_{m,n+1}+I_{m,n;m,n-1}P^+_{m,n-1}P_{m,n-1}) \qquad (1.2)$$

It should be pointed out that $J$ and $I$ denote interactions between bonds along cylinder axis while $I$ and $W$ denote interactions in discs.

Introducing notations

$$J_{m,n;m\pm 1,n}\equiv J; I_{m,n;m,n\pm 1}\equiv I \qquad (1.3)$$
$$Y_{m,n;m\pm 1,n}\equiv Y; W_{m,n;m,n\pm 1}\equiv W$$

one can reduce the Hamiltonian (1.2) to the form:

$$H=\sum_{m,n}(I+J)P^+_{m,n}P_{m,n}-\frac{1}{2}\sum_{m,n}\left(YP^+_{m,n}\left(P_{m+1,n}+P_{m-1,n}\right)+WP^+_{m,n}\left(P_{m,n+1}+P_{m,n-1}\right)\right)-$$

$$-\frac{1}{2}\sum_{n\nu}\left(JP^+_{m,n}P_{m,n}\left(P^+_{m+1,n}P_{m+1,n}+P^+_{m-1,n}P_{m-1,n}\right)+IP^+_{m,n}P_{m,n}\left(P^+_{m,n+1}P_{m,n+1}+P^+_{m,n-1}P_{m,n-1}\right)\right)$$

$$(1.4)$$

In order to formulate equations of motion for excitations in the system of hydrogen-bonds one has to find commutator of Pauli operator $P$ with Hamiltonian (1.4). Taking into account (1.1) and (1.4) one obtains:

$$[P_{m,n},H]=(J+I)P_{m,n}-\frac{1}{2}Y(P_{m+1,n}+P_{m-1,n})-\frac{1}{2}W(P_{m,n+1}+P_{m,n-1})$$

$$+YP^+_{m,n}P_{m,n}(P_{m+1,n}+P_{m-1,n})+WP^+_{m,n}P_{m,n}(P_{m,n+1}+P_{m,n-1})-$$

$$-J\left(P^+_{m+1,n}P_{m+1,n}+P^+_{m-1,n}P_{m-1,n}\right)P_{m\mu}-I\left(P^+_{m,n+1}P_{m,n+1}+P^+_{m,n-1}P_{m,n-1}\right)P_{m,n} \qquad (1.5)$$

To inquire further we shall go over to approximations which are equivalent to Tyblikov-s approuch in theory of magnetism [4,7]. The Tyblikov's approximation is applied to higher order Green's function. The equivalent of Tyblikov's approximation is going over from the commutators, by the use of approximation

$$P^+_{m,n}P_{m,n}\to <P^+P>=\frac{1-\sigma}{2} \qquad (1.6)$$

where $\sigma$ is ordering parameter of the system, to the approximate commutator

$$[P_{m,n},H]\approx \sigma(J+I)B_{m,n}-\frac{1}{2}Y(B_{m+1,n}+B_{m-1,n})-\frac{1}{2}W(B_{m,n+1}+B_{m,n-1}) \qquad (1.7)$$

In (1.7) Pauli operators $P$ are substituted by Bose operators $B$. This additional approximation with respect to the Tyblikov's one is in this case irrelevant, since our goal is finding one particle wave function of the system.

It should be pointed out that the mentioned approximation was succesfully applied in series of problems [8,9,10].

Taking into account the approximation used we can write:

$$[P_{m,n},H]\approx [B_{m,n},H] \qquad (1.8)$$

So the approximate equation of motion is:

$$i\hbar \dot{B}_{m,n} = EB_{m,n} = \sigma(J+I)B_{m,n} - \frac{1}{2}\sigma Y(B_{m+1,n} + B_{m-1,n}) - \frac{1}{2}\sigma W(B_{m,n+1} + B_{m,n-1})$$

(1.9)

where $E$ is energy of elementary excitations.

One particle wave funstion of the considered system is:

$$|\psi\rangle = \sum_{m,n} \mathcal{A}_{m,n} B^+_{m,n} |\psi\rangle \qquad (1.10)$$

From the condition $\langle\psi|\psi\rangle = 1$ it follows:

$$\sum_{m,n} |\mathcal{A}_{m,n}|^2 = 1 \qquad (1.11)$$

In order to find coefficients $\mathcal{A}_{m,n}$ one has to apply the operator

$$EB - [B_{m,n}, H] = 0$$

to wave function (1.10). In this way the system equations is obtained

$$\left\{[E - \sigma(J+I)]\mathcal{A}_{m,n} + \frac{1}{2}\sigma Y(\mathcal{A}_{m+1,n} + \mathcal{A}_{m-1,n}) + \frac{1}{2}\sigma W(\mathcal{A}_{m,n+1} + \mathcal{A}_{m,n-1})\right\} = 0 \qquad (1.12)$$

Since the quantum wire is translatory invariant along $z$-axis we can take the coefficients $\mathcal{A}_{m,n}$ in the form:

$$\mathcal{A}_{m,n} = F_n \, e^{imak} \qquad (1.13)$$

where $a$ is the lattice constant in $z$ direction and $k$ is the wave vector. After substitution (1.13) into (1.12) one obtains

$$F_{n+1} + F_{n-1} + \rho F_n = 0 \qquad (1.14)$$
$$n = 0, 1, 2, ..., N$$

where

$$\rho = \frac{E - [\sigma(J+I) - \sigma Y \cos ak]}{\frac{1}{2}\sigma W} \qquad (1.15)$$

Coefficients $F$ are related to disc molecules and consequently, they must satisfy the cyclicity condition:

$$F_{N+1+k} = F_k \qquad (1.16)$$
$$k = 0, \pm 1, \pm 2, ...$$

where $N$ is total number of molecules in one disc.

We shall look for the solution of (1.14) in the form

$$F_n = \Phi \cos n\varphi \qquad (1.17)$$

From (1.17) it follows:

$$F_{n+1} + F_{n-1} = F_n \, 2\cos\varphi \qquad (1.18)$$

This means that the system is satified if:

$$\rho = -2\cos\varphi \qquad (1.19)$$

The value of parameter $\varphi$ will be determined from the cyclicity condition by substituting (1.17) into (1.16). In such a way we obtain:

$$\sin\left(\frac{N+1}{2} + k\right)\varphi \, \sin\frac{N+1}{2}\varphi = 0$$

The last is satisfied if:

$$\varphi = \frac{2\pi\nu}{N+1} \tag{1.20}$$

$$\nu = 1,2,3,\ldots,N+1$$

Combining (1.20) and (1.15) we obtain the following formula for the energies of the elementary excitations micro tube

$$E_{k\nu} = \sigma(J+I) - \sigma Y\cos ak - \sigma W\cos\frac{2\pi\nu}{N+1} \tag{1.21}$$

The function $F$ is given by the formula:

$$F_n = \Phi \cos\frac{2\pi\nu}{N+1} n \tag{1.22}$$

while the cofficients $\mathcal{A}_{m,n}$ are given by

$$\mathcal{A}_{m,n} = \Phi \; e^{imak} \cos\frac{2\pi\nu}{N+1} n \tag{1.23}$$

The normalizing condition (1.11) has to be applied in two different ways. If $\frac{2\nu}{N+1}$ is integer the sum over index $n$ is equal $N+1$. If $\frac{2\nu}{N+1}$ is not integer, the sum over index $n$ is equal $\frac{N+1}{2}$. Consequently, we obtain double formula for wave function

$$|\psi_{k\nu}\rangle = \sum_{m,n} B^+_{m,n}|0\rangle \begin{cases} \sqrt{\dfrac{1}{M(N+1)}} e^{imak}\cos\dfrac{2\pi\nu}{N+1}n; \; \dfrac{2\nu}{N+1}\nu = \text{ceo broj} \\ \sqrt{\dfrac{2}{M(N+1)}} e^{imak}\cos\dfrac{2\pi\nu}{N+1}n; \; \dfrac{2\nu}{N+1}\nu \neq \text{ceo broj} \end{cases}$$

$$(1.24)$$

It is clear that the following is valid:

$$|\psi_{k\nu}\rangle = B^+_{k\nu}|0\rangle \tag{1.25}$$

Using this formula and (1.24) we can connect operators acting in momentum space $B^+_{k,\nu}$ with operators acting in configurational space $B^+_{m,n}$

$$B^+_{k,\nu} = B^+_{m,n} \begin{cases} \sqrt{\dfrac{1}{M(N+1)}} e^{imak}\cos\dfrac{2\pi\nu}{N+1}n; \; \dfrac{2\nu}{N+1} = \text{ceo broj} \\ \sqrt{\dfrac{2}{M(N+1)}} e^{imak}\cos\dfrac{2\pi\nu}{N+1}n; \; \dfrac{2\nu}{N+1} \neq \text{ceo broj} \end{cases} \tag{1.25}$$

Because of double normalizing condition the inverse transformation is doubled also:

$$B^+_{m,n} = B^+_{k,\nu} \begin{cases} \sqrt{\dfrac{1}{M(N+1)}} e^{imak}\cos\dfrac{2\pi\nu}{N+1}n; \; \dfrac{2n}{N+1} = \text{ceo broj} \\ \sqrt{\dfrac{2}{M(N+1)}} e^{imak}\cos\dfrac{2\pi\nu}{N+1}n; \; \dfrac{2n}{N+1} \neq \text{ceo broj} \end{cases} \tag{1.25}$$

and the function depending on $m$ and $n$ is the following:

$$|\psi_{mn}\rangle = \sum_{m,n} B^+_{k,\nu}|0\rangle \begin{cases} \sqrt{\dfrac{1}{M(N+1)}} e^{imak} \cos\dfrac{2\pi\nu}{N+1} n; \dfrac{2n}{N+1} = \text{ceo broj} \\ \sqrt{\dfrac{2}{M(N+1)}} e^{imak} \cos\dfrac{2\pi\nu}{N+1} n; \dfrac{2n}{N+1} \neq \text{ceo broj} \end{cases}$$

Taking into account the fact that the part of wave function corresponding to disc molecules is a real one it can be concluded that the probability current exists in z-direction, only,

$$j = \frac{\hbar}{2m^*i}\left(\langle\psi|\frac{\partial}{\partial(ma)}|\psi\rangle - \frac{\partial}{\partial(ma)}\langle\psi|\psi\rangle\right) = \frac{\hbar k}{m^*} \tag{1.28}$$

At the end of this part the two things will be noticed. The first of them is the fact that in the quantum tubes containing $2^n$ molecules in disc ($n \geq 2$), some specific selection rules take place. These rules prohibit appearance of some excitations on determining string of the tube. The same is valid for probability currence: some of them cannot flow along some of determining strings of the tube.

The second objection is rather of formal character. The usual approach in solving of the system of difference equations (1.18) is writing of this system in developed form and equating of determinant system with zero [11,12]. In the given case determinants are Fibonachy's type.

$$F_{N+1}(\rho) = \begin{vmatrix} \rho & 1 & 0 & 0 & 0 & \ldots & 0 & 0 & 1 \\ 1 & \rho & 1 & 0 & 0 & \ldots & 0 & 0 & 0 \\ 0 & 1 & \rho & 1 & 0 & \ldots & 0 & 0 & 0 \\ \ldots \\ 0 & 0 & 0 & 0 & \ldots & & 1 & \rho & 1 \\ 1 & 0 & 0 & 0 & \ldots & & 0 & 1 & \rho \end{vmatrix} \tag{1.29}$$

Equating to zero (1.29) one finds $\rho$ and $\varphi$. This calculation is sometimes complicated, as it was seen in this section. It is not necessary to look for the solution of Fibonachy equation. It is sufficient to analize cyclicity condition therefrom immediatelly follow the values of $\rho$ and $\varphi$.

## 2. Thermodynamics of quantum wire (fero- and paraelectric phase transitions)

Thermodynamical properties of quantum tubes will be analized by the method of two time Paulion Green's function.

$$\ll P_{m,n}(t)|P^+_{m',n'}(0)\gg = \Theta(t) < [P_{m,n}(t), P^+_{m',n'}(0)] > \tag{2.1}$$

In this formula $\Theta(t)$ is Hevisades step function:

$$\Theta(t) = \begin{cases} 1 & t > 0 \\ 0 & t < 0 \end{cases} \tag{2.2}$$

Derivating (2.1) over $t$ and using equations of motion of Pauli operator we come to the following equation:

$$i\hbar \frac{d}{dt} \ll P_{m,n}(t)|P^+_{m',n'}(0)\gg = i\hbar \delta_{m,m'}\delta_{n,n'}\delta(t) + \ll [P_{m,n}(t), H]|P^+_{m',n'}(0)\gg \tag{2.3}$$

After the substitution of commutator (1.5) into (1.3) we obtain:

$$i\hbar\frac{d}{dt}<<P_{m,n}(t)|P^+_{m',n'}(0)>> = i\hbar\sigma\delta_{m,m'}\delta_{n,n'}\delta(t) + (J+I)<<P_{m,n}(t)|P^+_{m',n'}(0)>> -$$

$$-\frac{1}{2}Y\left[<<P_{m+1,n}(t)|P^+_{m',n'}(0)>> + <<P_{m-1,n}(t)|P^+_{m',n'}(0)>>\right] -$$

$$-\frac{1}{2}W\left[<<P_{m,n+1}(t)|P^+_{m',n'}(0)>> + <<P_{m,n-1}(t)|P^+_{m',n'}(0)>>\right] +$$

$$+Y\left[<<P^+_{m,n}(t)P_{m,n}(t)P_{m+1,n}(t)|P^+_{m',n'}(0)>> + <<P^+_{m,n}(t)P_{m,n}(t)P_{m-1,n}(t)|P^+_{m',n'}(0)>>\right]$$

$$+W\left[<<P^+_{m,n}(t)P_{m,n}(t)P_{m,n+1}(t)|P^+_{m',n'}(0)>> + <<P^+_{m,n}(t)P_{m,n}(t)P_{m,n-1}(t)|P^+_{m',n'}(0)>>\right]$$

$$-J\left[<<P^+_{m+1,n}(t)P_{m+1,n}(t)P_{m,n}(t)|P^+_{m',n'}(0)>> + <<P^+_{m-1,n}(t)P_{m-1,n}(t)P_{m,n}(t)|P^+_{m',n'}(0)>>\right]$$

$$-I\left[<<P^+_{m,n+1}(t)P_{m,n+1}(t)P_{m,n}(t)|P^+_{m',n'}(0)>> + <<P^+_{m,n-1}(t)P_{m,n-1}(t)P_{m,n}(t)|P^+_{m',n'}(0)>>\right]$$

(2.4)

Applying to (2.4) Tyablikov's approximation which was described in the section one:

$$<<P^+_{m,n}(t)P_{m,n}(t)P_{a,b}(t)|P^+_{a',b'}(0)>> \approx \frac{1-\sigma}{2}<<P^+_{a,b}(t)|P^+_{a',b'}(0)>> \quad (2.5)$$

we reduce (2.4) to

$$i\hbar\frac{d}{dt}<<P_{m,n}(t)|P^+_{m',n'}(0)>> = i\hbar\sigma\delta_{m,m'}\delta_{n,n'}\delta(t) + \sigma(J+I)<<P_{m,n}(t)|P^+_{m',n'}(0)>> -$$

$$-\frac{\sigma}{2}Y\left[<<P_{m+1,n}(t)|P^+_{m',n'}(0)>> + <<P_{m-1,n}(t)|P^+_{m',n'}(0)>>\right] -$$

$$-\frac{\sigma}{2}W\left[<<P_{m,n+1}(t)|P^+_{m',n'}(0)>> + <<P_{m,n-1}(t)|P^+_{m',n'}(0)>>\right] + \quad (2.6)$$

After transformations

$$<<P_{m,n}(t)|P^+_{m',n'}(0)>> = \frac{1}{2\pi}\frac{1}{M}\frac{1}{N-1}\int_{-\infty}^{+\infty}d\omega\sum_{k,\varphi}<<P|P^+>>_{\omega,k,\varphi}e^{iak(m-m')+i\varphi(n-n')-i\omega t}$$

(2.7)

$$\varphi = \frac{2\pi\nu}{N+1} \quad \nu = 1,2,3,...,N+1,$$

we obtain the following equation for the Green's function

$$\left[E - \sigma(J+I) + \sigma Y\cos ak + \sigma W\cos\frac{2\pi\nu}{N+1}\right]\cdot G_{k,\nu}(\omega) = \frac{\sigma}{2\pi} \quad (2.8)$$

$$G_{k,\nu}(\omega) \equiv <<P|P^+>>_{\omega,k,\varphi}$$

Using spectral theorem for the Green's function [4,7] we come to the relation

$$\frac{1}{\sigma} = \frac{1}{\mathcal{N}}\sum\coth\frac{\sigma F_{k,\nu}}{2\Theta} \quad (2.9)$$

where

$$\mathcal{N} = M(N+1) \quad (2.10)$$

$$F_{k,\nu} = J + I - Y\cos ak - W\cos\frac{2\pi\nu}{N+1}$$

$$\Theta = k_B T$$

Since the ordering parameter σ is close to zero in the vicinity of transition temperature, the following approximation can be used:

$$\coth \varepsilon \approx \frac{1}{\varepsilon} + \frac{1}{3}\varepsilon \qquad (2.11)$$

Applying (2.11) to (2.9) we obtain

$$\sigma = \sqrt{L}\sqrt{1 - \frac{\Theta}{\Theta_c}} \qquad (2.12)$$

where

$$L = \frac{6\Theta_c M(N+1)}{\sum_{k,\nu} \frac{1}{F_{k,\nu}}} \qquad (2.13)$$

and

$$\Theta_c = \frac{1}{\frac{2}{M(N+1)} \sum_{k,\nu} \frac{1}{F_{k,\nu}}} \qquad (2.14)$$

Taking into account that $Y$ and $W$ are noticeably less than $J$ and $I$ one can write the function $\frac{1}{F}$ in (2.14) as geometric progression with exactness up to quadratic term. After simple calculations one can obtain the phase transition temperature:

$$\Theta_c = \frac{1}{2}\left(J + I - \frac{1}{2}\frac{Y^2 + W^2}{J+I}\right) \qquad (2.15)$$

The transition from ferroelectric to paraelectric phase is second kind phase transition since derivative of internal energy over temperature is infinite at the phase transition temperature.

The analizes of behaviour of quantum tube in paraelectric phase requires including of energy of external field $p\mathcal{E}$, where $p$ is dipole momentum.

$$\sigma = 1 - \frac{2\sigma}{M(N+1)}\sum_{k,\nu} \frac{1}{e^{\sigma\frac{p\mathcal{E}+J+I}{\Theta}} - 1} = \text{th}\left(\frac{p\mathcal{E}}{2\Theta} + \frac{J+I}{2\Theta}\right) \qquad (2.16)$$

In the analizes small terms $Y$ and $W$ will be neglected [13]. After these approximations and applications of the formula (2.9) it follows

$$\sigma \approx \text{th}\frac{p\mathcal{E}}{2\Theta} + \frac{\sigma(J+I)}{2\Theta}\frac{1}{\text{ch}^2 \frac{p\mathcal{E}}{2\Theta}}$$

wherefrom

$$\sigma\left[1 - \frac{J+I}{2\Theta}\frac{1}{\text{ch}^2 \frac{p\mathcal{E}}{2\Theta}}\right] = \text{th}\frac{p\mathcal{E}}{2\Theta} \qquad (2.17)$$

Since the electric susceptibility is defined as

$$\chi = \lim_{\varepsilon \to 0} \frac{\partial \sigma}{\partial \mathcal{E}} \qquad (2.18)$$

one has to derivate σ over field $\mathcal{E}$ tending $\mathcal{E}$ to zero. In this way one obtains

$$\chi = \frac{p}{2} \frac{1}{\Theta - \Theta_c^{(0)}} \quad (2.19)$$

where

$$\Theta_c^{(0)} = \frac{J+I}{2} \quad (2.20)$$

It is clear from (2.19) that electric susceptibility tends to infinity when the temperature of paraelectric phase tends to $\Theta_c^{(0)}$.

**Conclusion**

Authors made effort to give wide picture of cylindrical hydrogen bonded structures. The wave functions, energies of elementary excitations and phase transitions were investigated.

Finally it should be pointed out that the references which are quoted up to now were mainly used to explain some starting formulae.

Investigation of micro and nano cylindrical structures for the last several years are very popular so that citings of all attempts in this domain are practically impossible. We wish to note that the most of papers are dedicated to carbon nano tubes which possess good transfer properties (see for example [14-21]).